\documentclass{emulateapj}
\usepackage{apjfonts}

\slugcomment{Draft}
\slugcomment{\today}

\shorttitle{Star formation - galaxy mass relation in different environments}

\begin{document}

\title{Comparing the relation between star formation and galaxy mass in different environments}

\author{Benedetta Vulcani$^{1,2}$, Bianca M. Poggianti$^2$, Rose A.
Finn$^3$,  Gregory Rudnick$^4$, Vandana Desai$^5$, Steven Bamford$^6$}
\affil{$^1$Astronomical Department, Padova University, Italy,
$^2$INAF-Astronomical Observatory of Padova, Italy, $^3$Department of
Physics, Siena College, Loudonville, USA, $^4$University of Kansas, Department of Physics and Astronomy,
USA, $^5$Spitzer Science Center, California Institute of Technology, USA, $^6$School of Physics and Astronomy, University of Nottingham, United Kingdom
}

\begin{abstract}
Analyzing 24$\mu$m MIPS/Spitzer data and the [O II]3727 line
of a sample of galaxies at $0.4 \leq z \leq 0.8$ from the ESO
Distant Cluster Survey (EDisCS), we investigate the ongoing star
formation rate (SFR) and the specific star formation rate (SSFR) 
as a function of stellar mass in galaxy clusters and groups, 
and compare with field studies.
As for the field, we find a decline in SFR with time, indicating that
star formation (SF) was more active in the past, and a decline in SSFR as
galaxy stellar mass increases, showing that the current SF contributes
more to the fractional growth of low-mass galaxies than high-mass galaxies. 
However, we find a lower median SFR (by a factor of
$\sim$1.5) in cluster star-forming galaxies than in the field.
The difference is highly significant when all Spitzer and
emission-line galaxies are considered, regardless of color. It
remains significant at $z>0.6$ after removing 
red emission-line (REL) galaxies, 
to avoid possible AGN contamination.  
While there is overlap between the cluster and field SFR-Mass relations,
we find a population of cluster galaxies (10-25\%)
with reduced SFR for their mass. 
These are likely to be in transition from star-forming to passive.  
Comparing separately clusters and groups
at $z>0.6$, only cluster trends are significantly different from
the field, and the average cluster SFR at a given mass is $\sim2$ times
lower than the field. We conclude that the average SFR in
star-forming galaxies varies with galaxy environment at a fixed galaxy
mass.
\end{abstract}


\section{Introduction}
The SF activity, as other galaxy properties, varies
systematically with galaxy mass and redshift. Its trend as a function
of galaxy mass has been studied in the field at different epochs
(eg. \citealt{brinchmann04, noeske07a, elbaz07, daddi07, pannella09}).  
These works have found a
strong and rather tight correlation between current SFR and galaxy
stellar mass for field star-forming galaxies at all redshifts out to
$z=2$.  This correlation shifts to progressively higher SFRs at higher
$z$, maintaining remarkably its local slope.  These results suggest a
gradual decline in the SFR of most galaxies since z$\sim$1-2.

The SSFR, measuring the SFR per unit galaxy stellar mass, allows us to
study how the ongoing SF contributes to the mass growth for galaxies
of different masses, at different times. Lower-mass galaxies have
higher SSFRs than higher-mass galaxies \citep{feulner05,
perez05, zheng07, noeske07b},
supporting a scenario in which massive galaxies formed most of their
stars earlier and on shorter timescales, while less massive galaxies
evolve on longer timescales (``downsizing'').

There are several reasons to expect that the SFR-Mass and SSFR-Mass relations 
should depend on environment. While
 fast-acting environmental effects are unlikely to influence the SFR-Mass
 relationship of star-forming galaxies, any physical mechanism slowly
 affecting the amount of gas available for SF should result in a
 slowly declining SFR, therefore a different SFR-Mass relation
 with environment.  Examples are the loss of halo gas reservoir
 included in hierarchical galaxy formation models (``strangulation'',
 \citealt{larson80, font08, mccarthy08}), and the interruption of cold
 gas streams in dense environments that would leave un-fueled galaxies to
 slowly consume their disk gas \citep{keres05}.

In contrast, several works have reported a lack of
environmental dependence of the distribution on current SF
activity as measured both from optical lines and infrared emission
(\citealt{balogh04a, rines05, bai06, bai07, bai09}, but see \citealt{wolf09}), and others have
failed to identify large population of galaxies in ``transition'' from
the red sequence to the blue cloud in dense environments \citep{balogh04b,
weinmann06, cassata07}.

However, the relations between SFR and SSFR with mass have not yet
been studied in groups and clusters, which should be the most
direct way to clearly discriminate between mass and environmental
trends. Should the SFR-Mass relation be universal, the evolution
of the red galaxy fraction would have a galaxy intrinsic origin,
and environmental effects such as strangulation 
could not be relevant. 

In this paper we make a first attempt to investigate this issue by
studying galaxies in clusters and groups at intermediate redshifts
($0.4<z<0.8$) using the ESO Distant Cluster
Survey (EDisCS) dataset and comparing with field galaxies at similar redshifts.

We adopt ($H_0$, ${\Omega}_m$,
${\Omega}_{\lambda}$) = (70 $\rm km \, \, s^{-1} \, Mpc^{-1}$, 
0.3, 0.7). Values of $M_{\ast}$ and SFR
are based on the \cite{salpeter55} Initial Mass Function, in the range
of mass 0.1-125 $M_{\odot}$.

\section{DATA SET}
In our analysis, we use 604 spectroscopically confirmed EDisCS members of 16
clusters with velocity dispersions $\sigma > 400 \rm \, km \, s^{-1}$
and 10 groups ($ 150 < \sigma < 400 \rm \, km \, s^{-1}$) as in
\cite{poggianti09} (see also \citealt{halliday04}, 
\citealt{milvang-jensen08}). In the following, we refer to
clusters and groups collectively as ``clusters'', unless otherwise
stated.

Ours is effectively
an I-band selected sample with high quality multiband optical and
near-IR photometry \citep{white05} and spectroscopy (\citealt{halliday04}, 
\citealt{milvang-jensen08}), with a 97\% spectroscopic
success rate (number of redshifts/number of spectra)
at the magnitudes used here.

We estimate galaxy stellar masses using photo-z fitting total
absolute magnitudes \citep{pello09} and the relation 
between mass-to-light $M/L_B$ ratio and rest-frame
($B-V$) color for solar metallicity from \cite{bell01}, 
($ \log \frac{M}{L_B}= -0.51+1.45(B-V)$). 
The internal accuracy of the measured masses 
is $\sim$0.15{\it dex}. 
The spectroscopic magnitude limit 
($I=23$ at $z=0.8$ and $I=22$ at $z=0.6$) corresponds to a mass
limit  $\log M_{\ast}= 10.8$ $M_{\odot}$
for galaxies of all colors, 
and $\log M_{\ast}= 10.5$ $M_{\odot}$ for blue galaxies (see below for 
our red/blue limit).

To estimate SFRs,
we use the IR luminosities of \cite{finn09}.  
The IR luminosities are derived from {\it Spitzer} 24$\mu$m imaging,
and the observed 24$\mu$m flux
is converted to total IR luminosity using the models of \cite{dale02}.  
The IR luminosity is converted to $SFR_{IR}$ according to
\cite{kennicutt98}:
$SFR_{IR}(M_{\odot} \, yr^{-1}) =4.5\times 10^{-44} \times L_{IR}(\rm ergs \, s^{-1})$,
assuming that the mid-IR emission of the great majority of 
distant cluster galaxies is  
powered by starbursts rather than AGNs, as found by previous studies 
(Finn et al. 2009 and references therein).
The median IR luminosity error is 7\%, and is always less than 23\%.
The $SFR_{IR}$ error associated with estimating the IR luminosity
from the observed 24$\mu$m  flux ranges from 5 to 22\%, depending on the
cluster redshift (Finn et al. 2009).
The 80\% completeness limit of our Spitzer data corresponds
to a $SFR_{IR}$ of \mbox{$\sim$10.3 $M_{\odot} yr^{-1}$} at $z=0.6$
and $\sim$13 $M_{\odot} \, yr^{-1}$ at $z=0.8$.

We also use the $SFR_{[OII]}$ \citep{poggianti08} from the observed
[OII] luminosity using the \cite{kewley04} conversion: $SFR_{[OII]}
(M_{\odot} \, yr^{-1}) =1.26\times 10^{41} L_{[OII]}(\rm ergs \,
s^{-1})$, corrected to our IMF. The $SFR_{[OII]}$ detection limit is
$\sim$0.3$M_{\odot} \, yr^{-1}$. The EDisCS [OII] detections of even
weak lines are very robust, having been confirmed by manual inspection
of all 2D spectra.  The $SFR_{[OII]}$ errors in $SFR_{[OII]}$ range from
$\sim$0.05{\it dex} to $\sim$0.4{\it dex}, 
with a mean error of $\sim$0.1{\it dex}.

To account for both obscured and
unobscured SF, in galaxies with a 24$\mu$m
detection (all of which have emission-lines in their spectra),
we use the total
$SFR_{tot}=SFR_{IR}+SFR_{[OII]}$, without correcting the 
[OII] estimate for dust extinction.
For galaxies without a 24$\mu$m detection, we use the $SFR_{[OII]}$
corrected for dust.
The [OII] extinction correction is estimated from the 
correlation between the uncorrected $SFR_{[OII]}$
and $E(B-V)$ 
observed at low-z:
$E(B-V)=0.165 \log(SFR_{[OII]}) +0.315$
(\citealt{fritz09}, \citealt{kewley04}).
This $E(B-V)$ is derived from the Balmer decrement, thus is
appropriate for emission-lines. We
adopt the \cite{sudzius96} 
Galactic mean interstellar extinction law for which
$\frac{A_{3727}}{E(B-V)}=4.749 $.
Using either the total 24$\micron$+[OII] or the dust corrected [OII] SFRs,
the SSFR is simply $SSFR = \frac{SFR}{M_{\ast}}$.

Galaxies without a 24$\mu$m detection are further divided into
red and blue, in order to separate those that can be assumed
to be powered by SF (blue) from those that could
be strongly contaminated by an AGN (red).
Following \cite{noeske07a},
the color separation
is defined by \cite{willmer06}:
$$
(U-B)_{rest-frame} \geq -0.032(M_B+21.52)+0.454-0.25.
$$ 

Based on visual morphological classifications by
\cite{desai07}, blue emission-line (BEL) galaxies and
Spitzer-detected galaxies have mostly late-type morphologies
($\sim 75\%$ and $\sim 95\%$, respectively), in agreement with 
the assumption that they are star-forming.

In addition, a fraction of the REL galaxies could have their [OII] 
powered by a residual low level of SF,
instead of being dominated by an AGN.

In our dataset, we currently do not have a way to quantify AGN contamination.
External estimates can vary significantly: 
at low-z, \cite{yan06} in their SDSS sample found that only 
$\sim 10 \%$ of red galaxies with [OII] in emission are
characterized by SF.
At higher redshifts, \cite{noeske07a} find a higher
fraction of probable star formation-dominated REL
systems, up to 30\%. In broad agreement with this, we find that
32\% of our REL galaxies  have late-type
morphologies, while only 50\% are ellipticals. This suggests that
at least a third of galaxies in this class are indeed star-forming.
Moreover, analyzing the EDisCS optical spectra, Sanchez-Blazquez et al. (2009) 
conclude that most of our REL galaxies are dusty and 
star-forming. This agrees with the large population of red star-forming cluster galaxies
identified at $z\sim0.2$ \citep{wolf09}.

However, since we cannot be certain of exactly how many red objects should be
considered star-forming, in our analysis we analyze two different
cases: in the first case we assume that the [OII] emission in {\it
all} red galaxies is dominated by an AGN, and we exclude them from our
analysis. In the second case, we include REL
galaxies. These cases should bracket the real situation occurring in
nature.

\section{COMPARISON WITH THE FIELD}
We compare our results to those for the field at $0.4\leq z \leq 0.8$ from
\cite{noeske07a,noeske07b}, who studied the SFR and the SSFR in field galaxies
from the All-Wavelength Extended Groth Strip International Survey 
(AEGIS).\footnote{The Kroupa values in their paper can 
be transformed to a Salpeter IMF multiplying by a factor $2$ 
(Noeske 2009, private communication).}  

Our method to derive SFRs is similar to theirs.  For galaxies with
robust 24$\mu$m detections, they determined the total SFR by summing
the SFR derived from 24$\micron$ data with that derived from the
emission lines uncorrected for extinction, as we do.  For galaxies
below the 24$\mu$m detection limit, they estimated
extinction-corrected SFRs from emission line fluxes using the observed
average Balmer decrement, for a fixed $A_{\rm
H\alpha}=1.30$ value.  This method overestimates the extinction in
galaxies with low SFRs \citep{noeske07a}. To avoid systematic
effects due to a different dust treatment for field and cluster galaxies,
we de-corrected the AEGIS emission-line data using their value of extinction,
then we applied our own method of dust
correction to their datapoints. 

Our comparison with the field is meaningful
only for those galaxies with mass and SFR 
above the highest between our and AEGIS limits, where
we are sure both samples are unbiased. 
These limits are: a) at $z<0.6$,  
$M_{\ast} \geq 10^{10.8} M_{\odot}$ 
when we consider both blue and red galaxies 
($10^{10.5} M_{\odot}$
when we do not consider red galaxies), 
and $SFR_{[OII]_{corr}} \geq 1.2 M_{\odot} yr^{-1}$;
b) at $z>0.6$, $M_{\ast} \geq 10^{10.8} M_{\odot}$ and 
$SFR_{[OII]_{corr}} \geq 1.65 M_{\odot} \, yr^{-1}$.
Galaxies above our mass limit but
below the $SFR_{[OII]}$ limits will be disregarded
in our analysis, as they make a negligible
contribution to the SFR 
census.
This leaves us with a final 
cluster sample of 127 galaxies, and a field sample of 426 galaxies.

\section{RESULTS}

\begin{figure*}[!h]
\begin{minipage}[c]{175pt}
\centering
\includegraphics[scale=0.45,clip = false, trim = 0pt 150pt 0pt 0pt]
{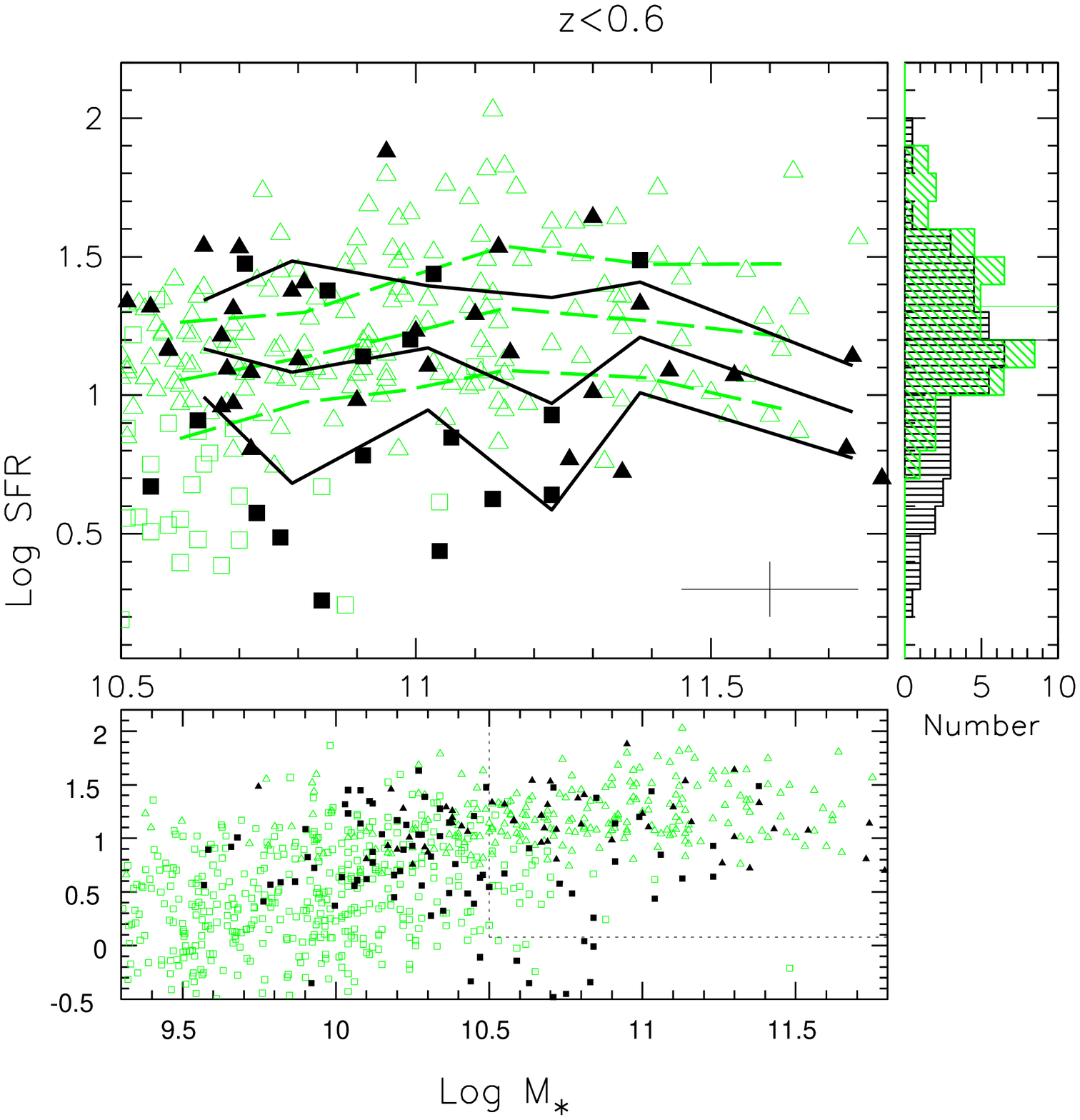}
\vspace*{2.2cm}
\end{minipage}
\hspace{2.5cm}
\begin{minipage}[c]{175pt}
\centering
\includegraphics[scale=0.45,clip = false, trim = 0pt 150pt 0pt 0pt]
{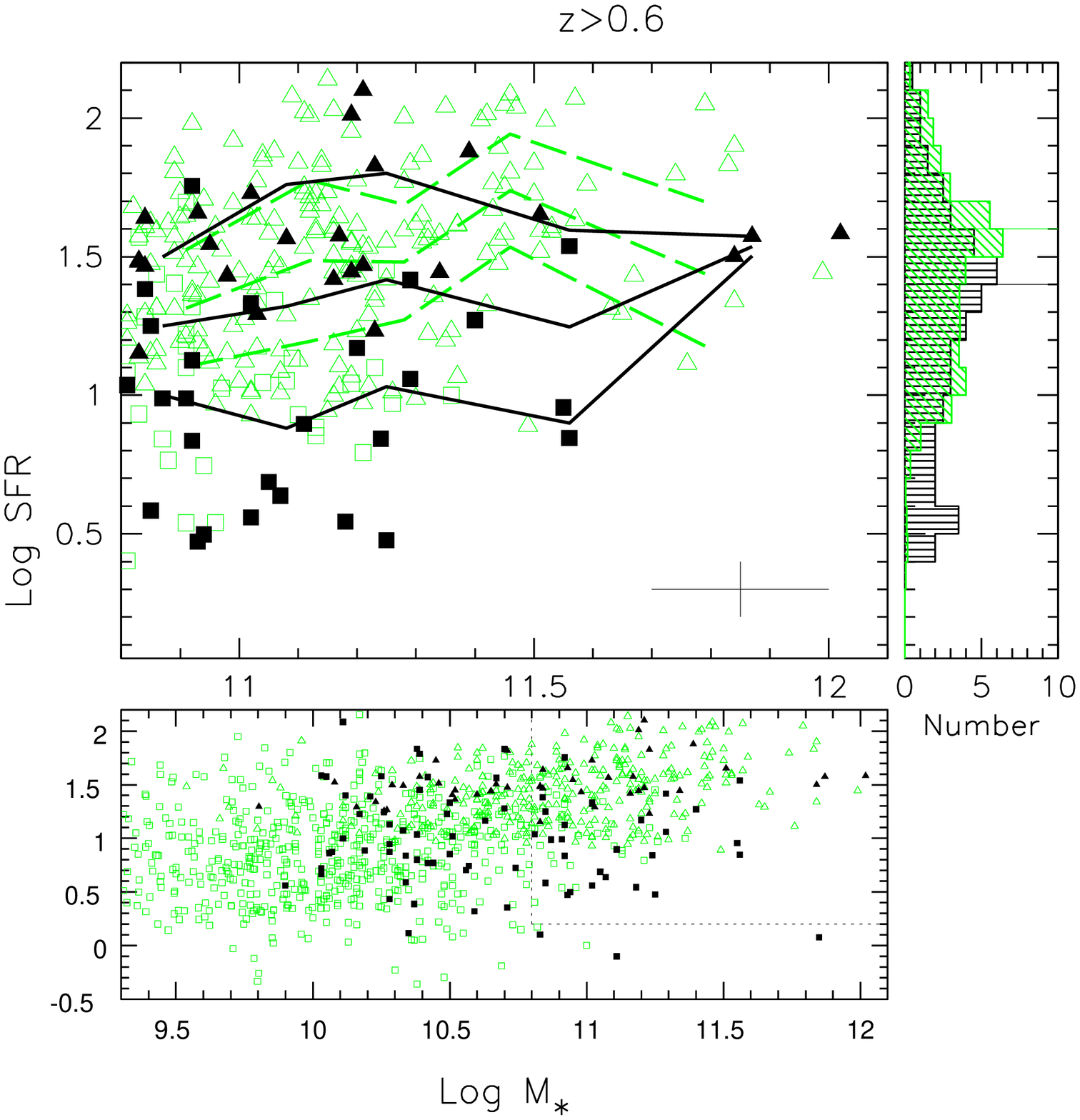}
\vspace*{2.2cm}
\end{minipage}
\hspace{1cm}
\caption{SFR--M$_{\ast}$ relation of cluster and field
galaxies at $z<0.6$ ({\bf left panel}) and $z>0.6$ ({\bf right panel}). 
Only 24$\mu m$+BEL galaxies are plotted. 
{\bf Upper left window}: only data points above the completeness limits.
{\bf Bottom window}: all galaxies.
Black filled symbols: cluster galaxies (EDisCS).
Green empty symbols: field galaxies (AEGIS). 
Triangles: combined SFRs from MIPS 24$\mu m$ and emission lines.
Squares: [OII] dust-corrected SFRs.
In the upper window, lines represent the median 
and the 25 and 75 percentiles
for clusters (solid black) and field (dashed green). 
Typical EDisCS errorbars are in the bottom-right.  
In the bottom window, lines show the mass and SFR limits.
{\bf Upper right window}: 
SFR distribution of galaxies above the completeness limits, 
selecting the same
mass distribution in clusters and field, for the
average of the 1000 simulations. The number of field galaxies
is normalized to the number of cluster galaxies. 
Black horizontal histogram: EDisCS. Green slanted histogram: AEGIS. 
Black (EDisCS) and green (AEGIS) solid lines
are the mean values of the distributions.
\label{SFR}}
\end{figure*}

\begin{figure*}[!h]
\begin{minipage}[c]{175pt}
\centering
\includegraphics[scale=0.45,clip = false, trim = 0pt 150pt 0pt 0pt]
{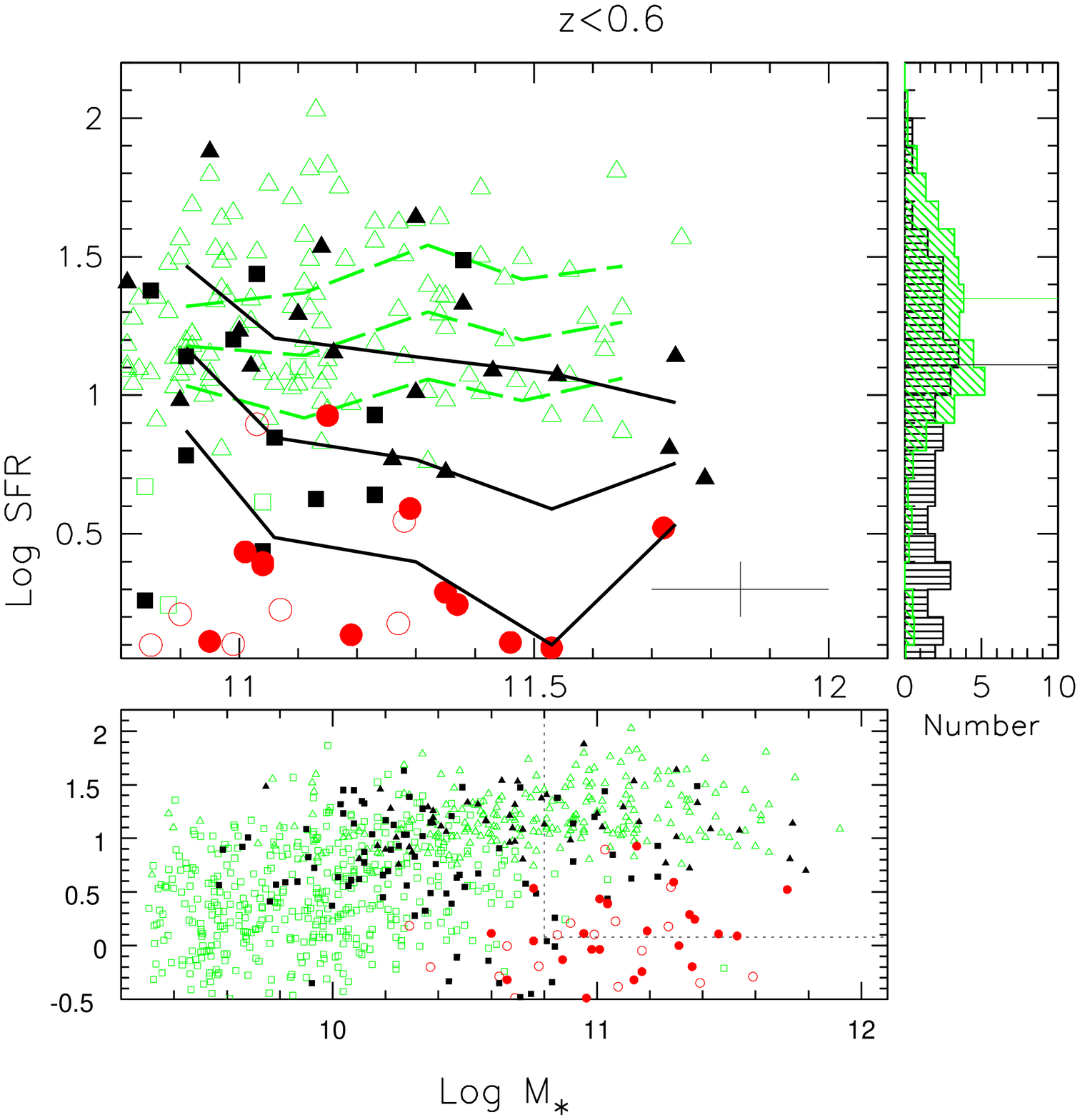}
\vspace*{2.2cm}
\end{minipage}
\hspace{2.5cm}
\begin{minipage}[c]{175pt}
\centering
\includegraphics[scale=0.45,clip = false, trim = 0pt 150pt 0pt 0pt]
{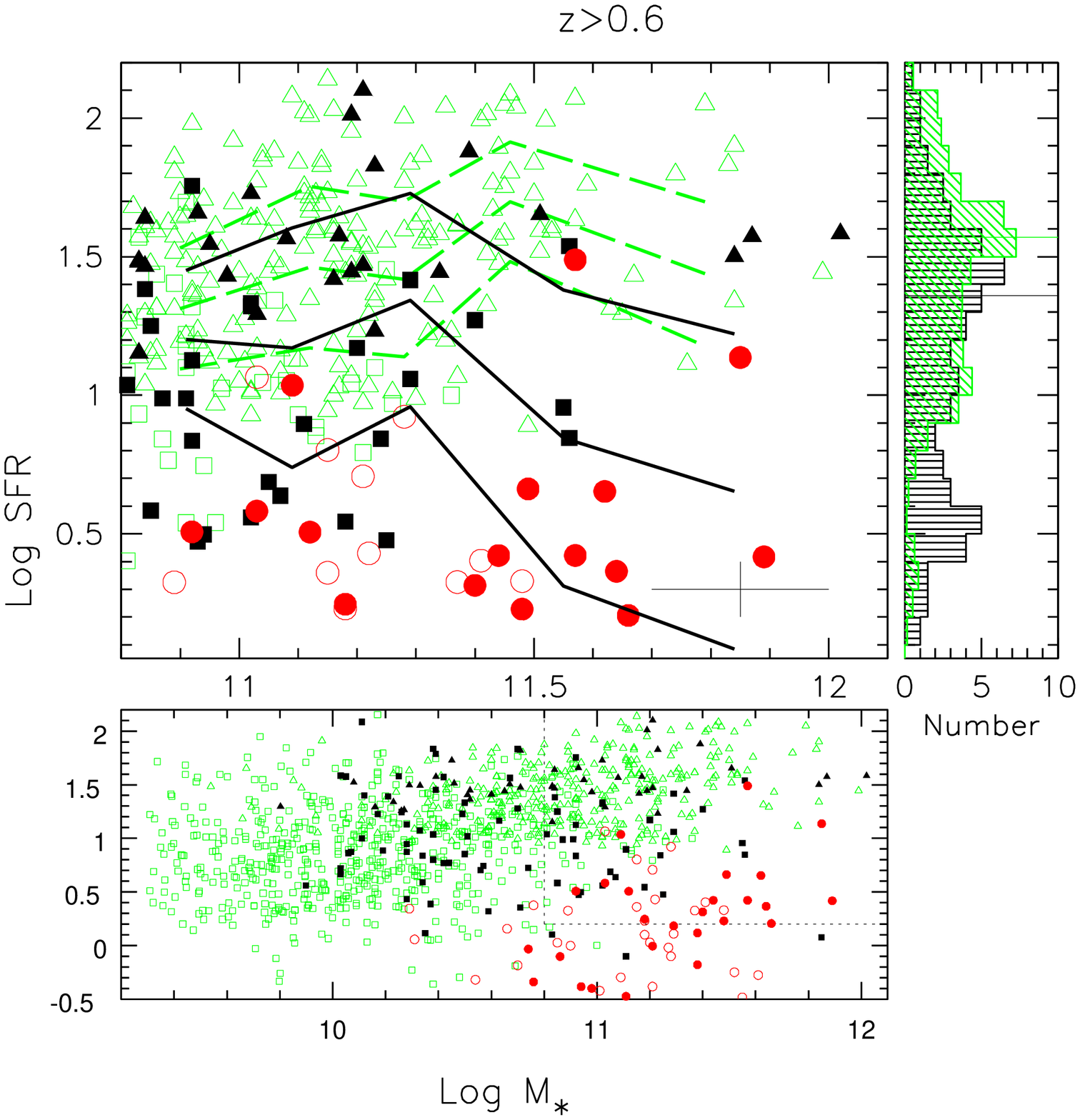}
\vspace*{2.2cm}
\end{minipage}
\hspace{1cm}
\caption{SFR--M$_{\ast}$ relation of cluster and field
galaxies when REL galaxies are included. 
Red filled circles: cluster REL galaxies. 
Red empty  circles: field REL galaxies. 
Other symbols as in Fig.~\ref{SFR}.  
REL galaxies are included
in the histograms.
 \label{SFR2}}
\end{figure*}

We show the SFR-Mass relation in different environments 
in Fig.~\ref{SFR}, where only 24$\micron$ and BEL
galaxies are considered star-forming.  We calculate the median values
of SFR and of SSFR, 
and the 25 and 75 percentiles.

From Fig.~\ref{SFR}, we note a change with redshift, in both the field
and clusters, as the average 
SFRs at $z\sim 0.7$ are shifted to higher values
compared to $z \sim 0.5$ at the same mass. 

The main result of Fig.~\ref{SFR} is that, at both redshifts, cluster
SFRs are on average systematically lower than field SFRs at the
same mass. Clusters have in general lower median SFRs than the field.
This is due to a population of cluster
galaxies lying below the field 25 percentile, that 
represent 34\% of the whole cluster population, thus
a $\sim 10$\% excess of galaxies with ``reduced'' SFR for their mass.

To avoid the influence of the mass
distribution, we performed 1000 Monte Carlo simulations
extracting randomly from the field sample a subsample
with the same mass distribution as the clusters.  
The SFR distributions are shown as histograms in Fig.~\ref{SFR}.
At $z<0.6$, due to poor number statistics\footnote{Doubling the number 
of galaxies, differences become significant at $>$95\% in 80\% of the cases.}, 
a Kolmogorov-Smirnov test cannot 
reject the null hypothesis of similar
SFR distributions in clusters and field, finding 
a probability less than 90\% in 54\% of the cases and a 
probability $>95\%$ in 29\% of the simulations.
At $z>0.6$, the K-S test rejects the hypothesis of similar cluster
and field distributions 
with a $>$95\% probability in 87\% of the simulations.

We note that, although a correlation is evident when considering 
galaxies over a wide mass range (see bottom windows in Fig.~\ref{SFR}),
the SFR-Mass relation is flat
above our mass limit. A Spearman test yields
a significant positive correlation only for the field at $z>0.6$ (99.9\%),
and no correlation in all other cases.

In Fig. \ref{SFR2}, we show the results considering also
REL galaxies as star-forming. We recall that
at least for some of them 
the [OII] emission likely arises
from ongoing SF. Now the difference
between field and clusters is more striking, and 
becomes progressively more pronounced towards more massive galaxies.
50\% of the whole cluster population 
have SFRs below the field 25 percentile, therefore $\sim25$\% have
``reduced'' SFRs for their mass compared to the field.
The K-S test on mass-matched cluster and field samples
rejects the null hypothesis of similarity between the
two environments with a probability of 100\% ($z<0.6$) and $>95\%$ ($z>0.6$)
 in all the simulations 
(see histograms in Fig. \ref{SFR2}).
As before, no SFR-Mass correlation is detected by a Spearman test, except
for the field at $z>0.6$ at 99.9\%.

Our results highlight a
change in the SFR-Mass relation with environment. 
To quantify this change, we compute the mean SFR in 
our mass-matched simulations, therefore removing the effects
of different mass distributions. 
Including both redshift bins, for the sample of 24$\micron$+BEL galaxies 
the mean SFR in clusters is 1.35$\pm$0.15 times lower than in the field.
Including REL galaxies, it is 1.63$\pm$0.20 times lower
than the field.\footnote{Errors are computed as
bootstrap standard deviations.}

\begin{figure*}[!h]
\begin{minipage}[c]{175pt}
\centering
\includegraphics[scale=0.45,clip = false, trim = 0pt 150pt 0pt 0pt]
{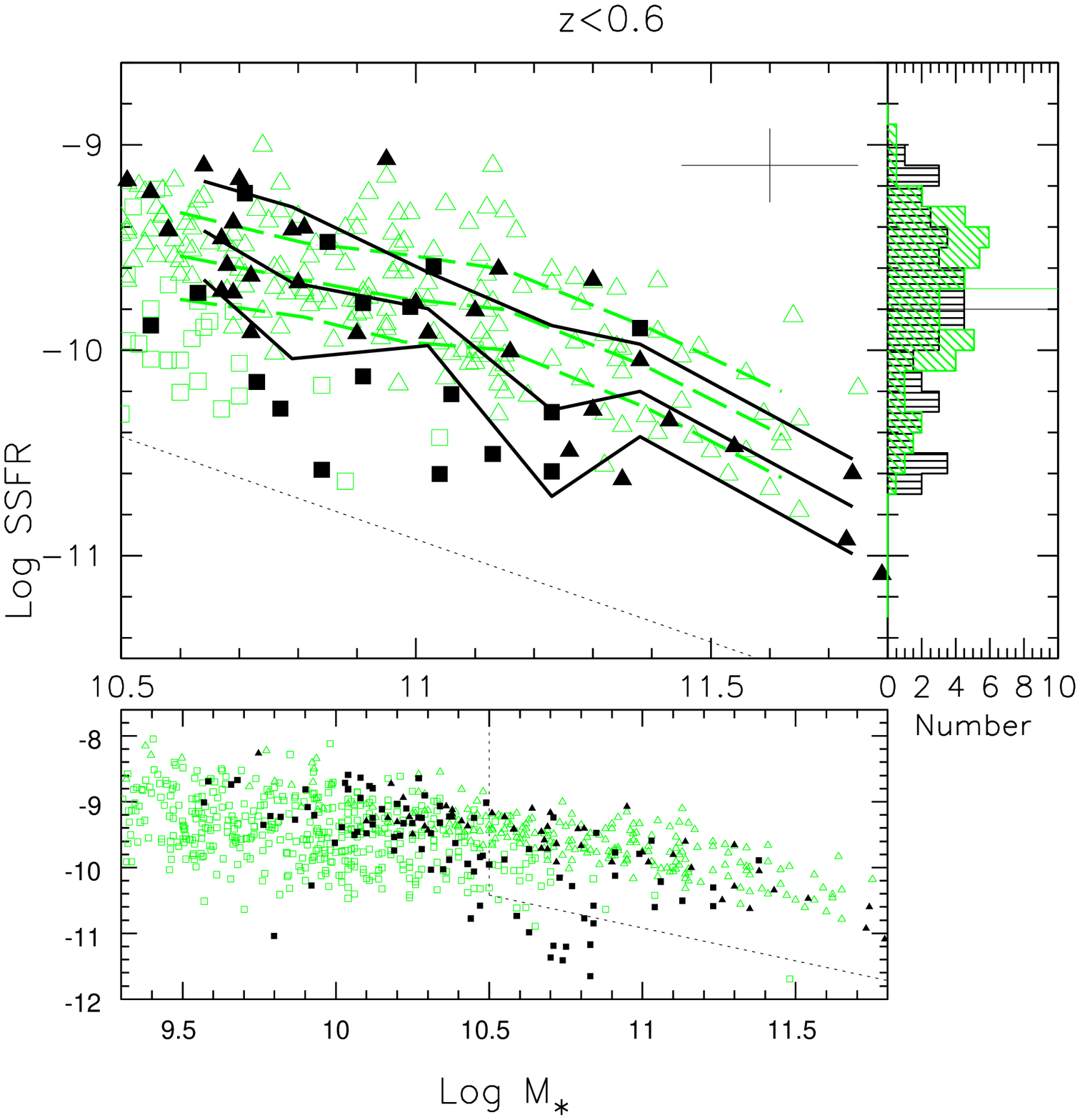}
\vspace*{2.2cm}
\end{minipage}
\hspace{2.5cm}
\begin{minipage}[c]{175pt}
\centering
\includegraphics[scale=0.45,clip = false, trim = 0pt 150pt 0pt 0pt]
{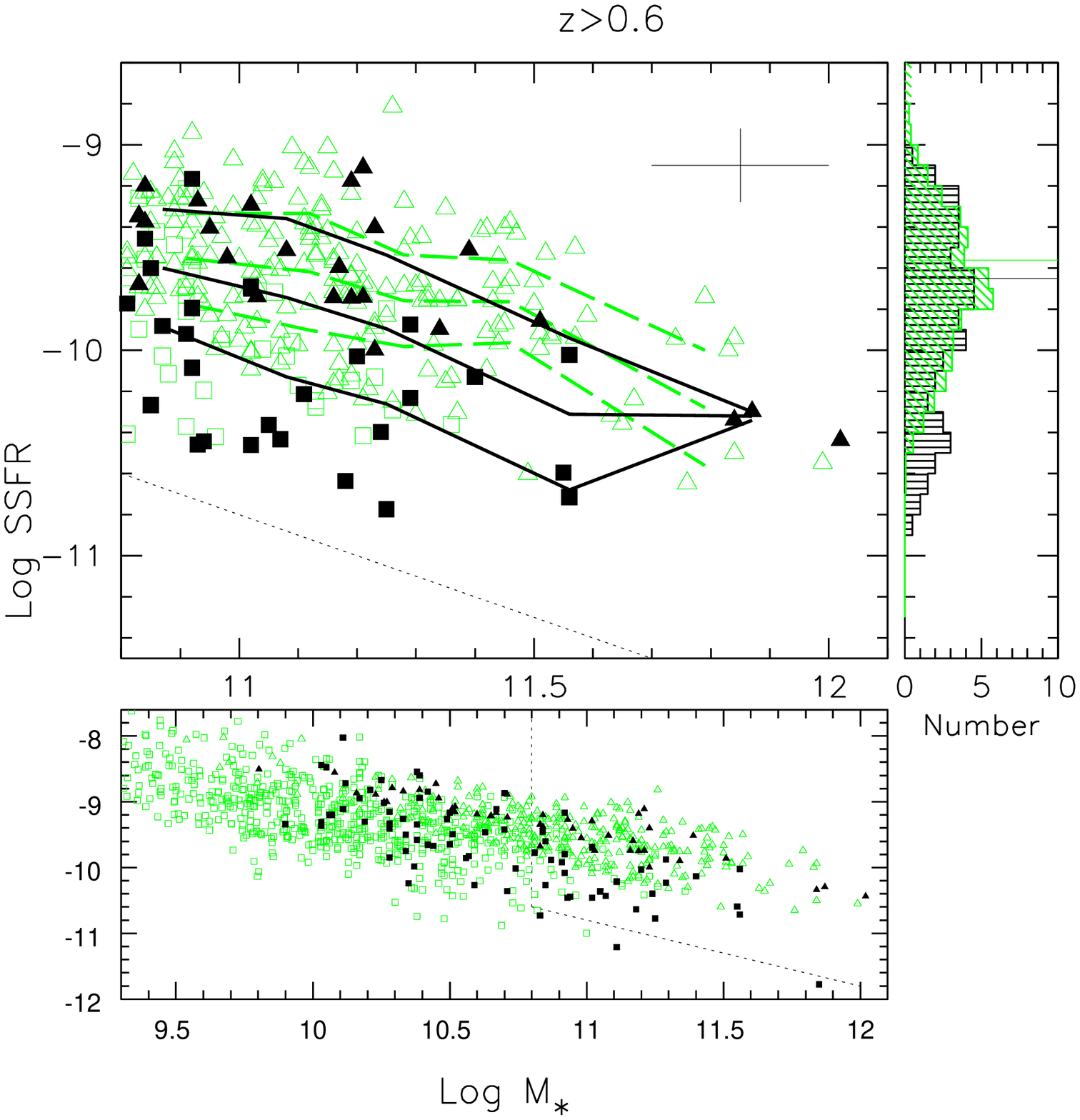}
\vspace*{2.2cm}
\end{minipage}
\hspace{2.5cm}
\caption{SSFR--M$_{\ast}$ relation of cluster and field
galaxies. Only 24$\mu m$+BEL galaxies are plotted.
The dotted lines mark the completeness limits.
Symbols are as in Fig.~\ref{SFR}. Typical EDisCS 
errorbars are shown in 
the top-right corner. \label{SSFR} }
\end{figure*}

\begin{figure*}[!h]
\begin{minipage}[c]{175pt}
\centering
\includegraphics[scale=0.45,clip = false, trim = 0pt 150pt 0pt 0pt]
{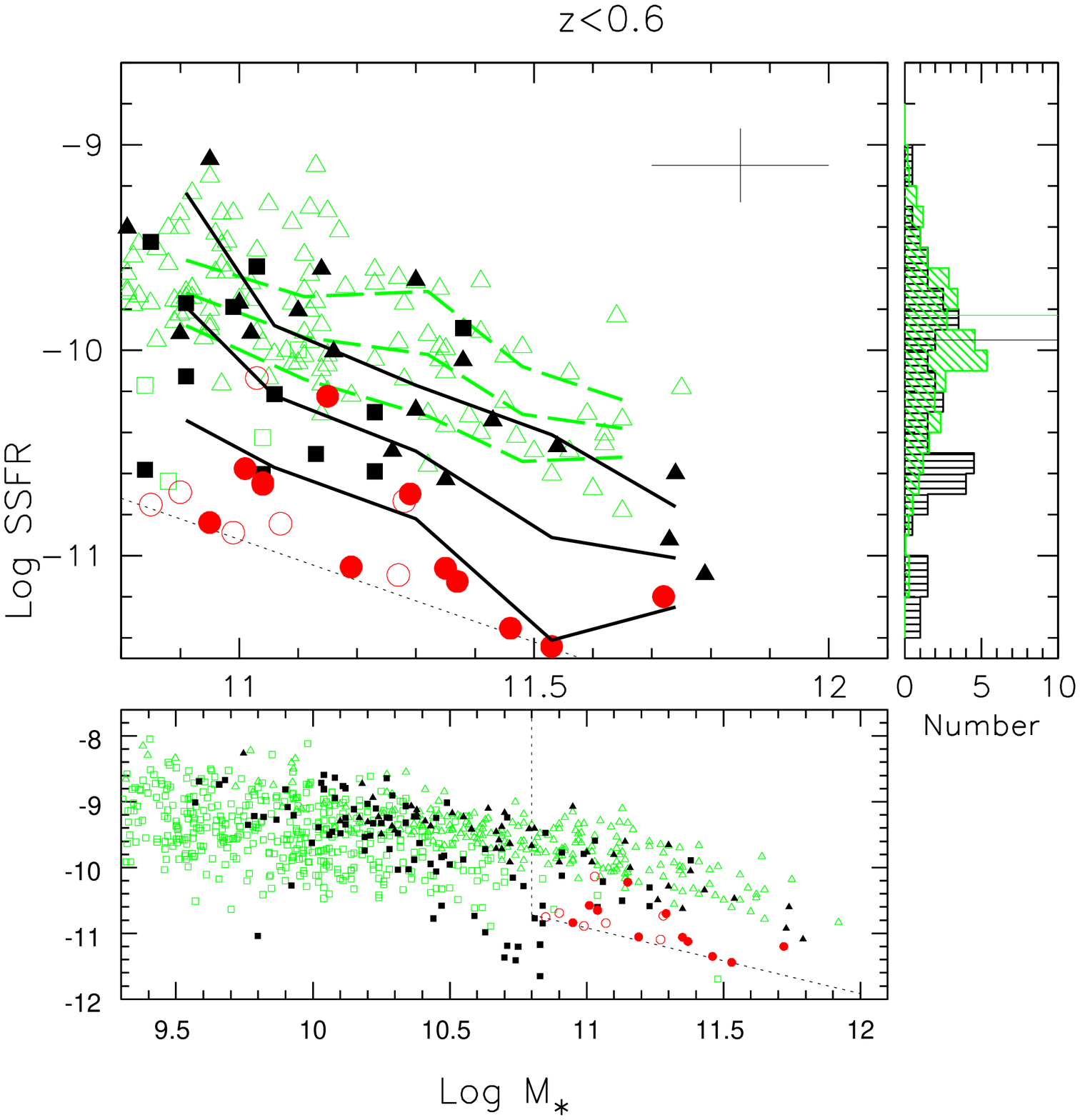}
\vspace*{2.2cm}
\end{minipage}
\hspace{2.5cm}
\begin{minipage}[c]{175pt}
\centering
\includegraphics[scale=0.45,clip = false, trim = 0pt 150pt 0pt 0pt]
{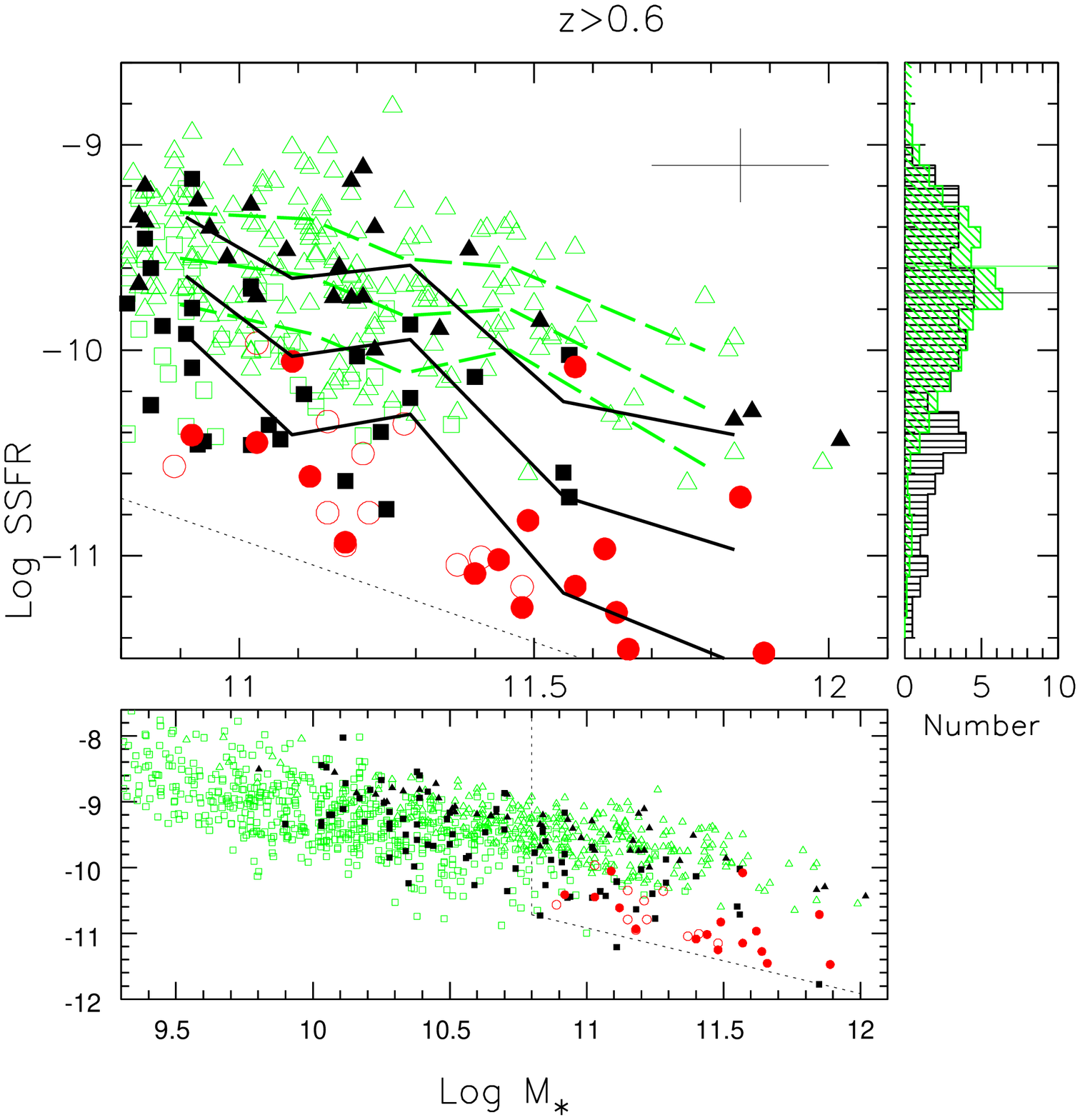}
\vspace*{2.2cm}
\end{minipage}
\hspace{2.5cm}
\caption{SSFR--M$_{\ast}$ relation of cluster and field
galaxies. Galaxies with 24$\mu m$
detections and all emission line galaxies are plotted,
regardless of (U-B) color. 
The dotted lines mark the completeness limits.
Symbols are as 
in Fig.~\ref{SFR2}. \label{SSFR2} }
\end{figure*}

In Figs.~\ref{SSFR} and \ref{SSFR2}, we show the results for the
SSFR-Mass relation.  Cluster and field galaxies follow a qualitatively
similar decreasing trend of SSFR with mass (Spearman anticorrelation
probability always $>99.9\%$), but cluster galaxies tend to have a
lower SSFR than field galaxies of similar mass, as expected from the
results discussed above.  A K-S test confirms this when red galaxies
are included (probabilities always $>95\%$). For 24$\micron$+BEL
galaxies, the differences are not statistically significant, with the
K-S test giving a probability $>95\%$ only in 53\% of the simulations
at $z>0.6$, and only in 4\% of simulations at $z<0.6$.  Including both
redshift bins, the ratio of average SSFR between field and clusters
above our mass limit is 1.20$\pm$0.14 for 24$\micron$+BEL galaxies,
and 1.31$\pm$0.17 for all galaxies.

This result shows that, in all environments, the mass growth rate at a
given mass decreases with time (cf. left and right panel of
Figs.~\ref{SSFR} and \ref{SSFR2}) and it is lower for higher mass
galaxies (SSFR and mass are anticorrelated).  However, a fraction 
of the star-forming cluster galaxies are building-up their stellar mass
at a lower rate than field galaxies: 10\% and 30\% of
24$\micron$+BEL and all galaxies, respectively, lie below the field
25 percentile.

Moreover, the cluster trends are steeper than the field trends
(best-fit slopes differ by $>1\sigma$),
again suggesting that SF in more
massive galaxies differs more strongly with environment than SF 
in lower mass galaxies.

\subsection{Clusters versus groups}
\begin{figure*}[!h]
\begin{minipage}[c]{175pt}
\centering
\includegraphics[scale=0.45,clip = false, trim = 0pt 150pt 0pt 0pt]
{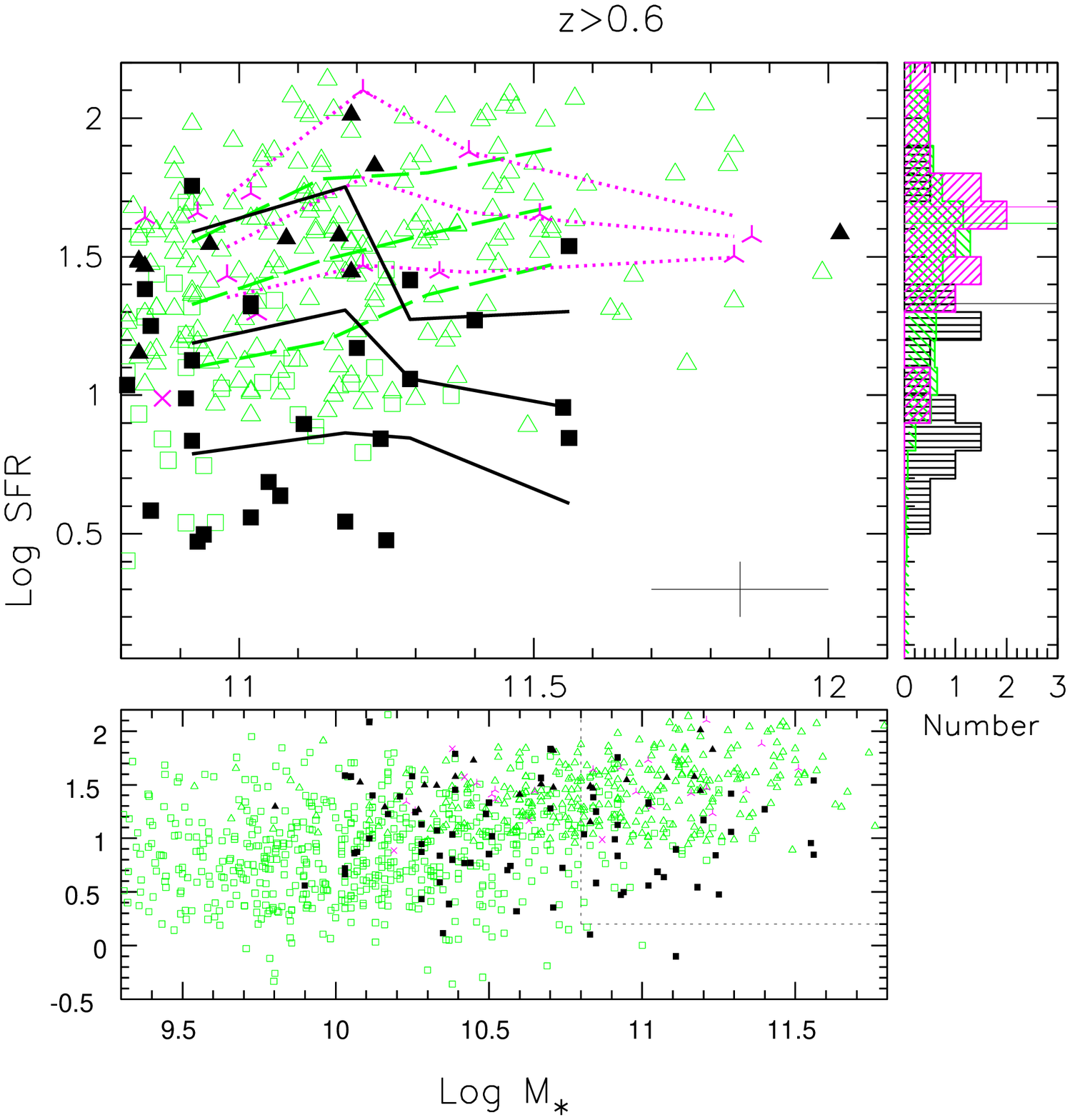}
\vspace*{2.2cm}
\end{minipage}
\hspace{2.5cm}
\begin{minipage}[c]{175pt}
\centering
\includegraphics[scale=0.45,clip = false, trim = 0pt 150pt 0pt 0pt]
{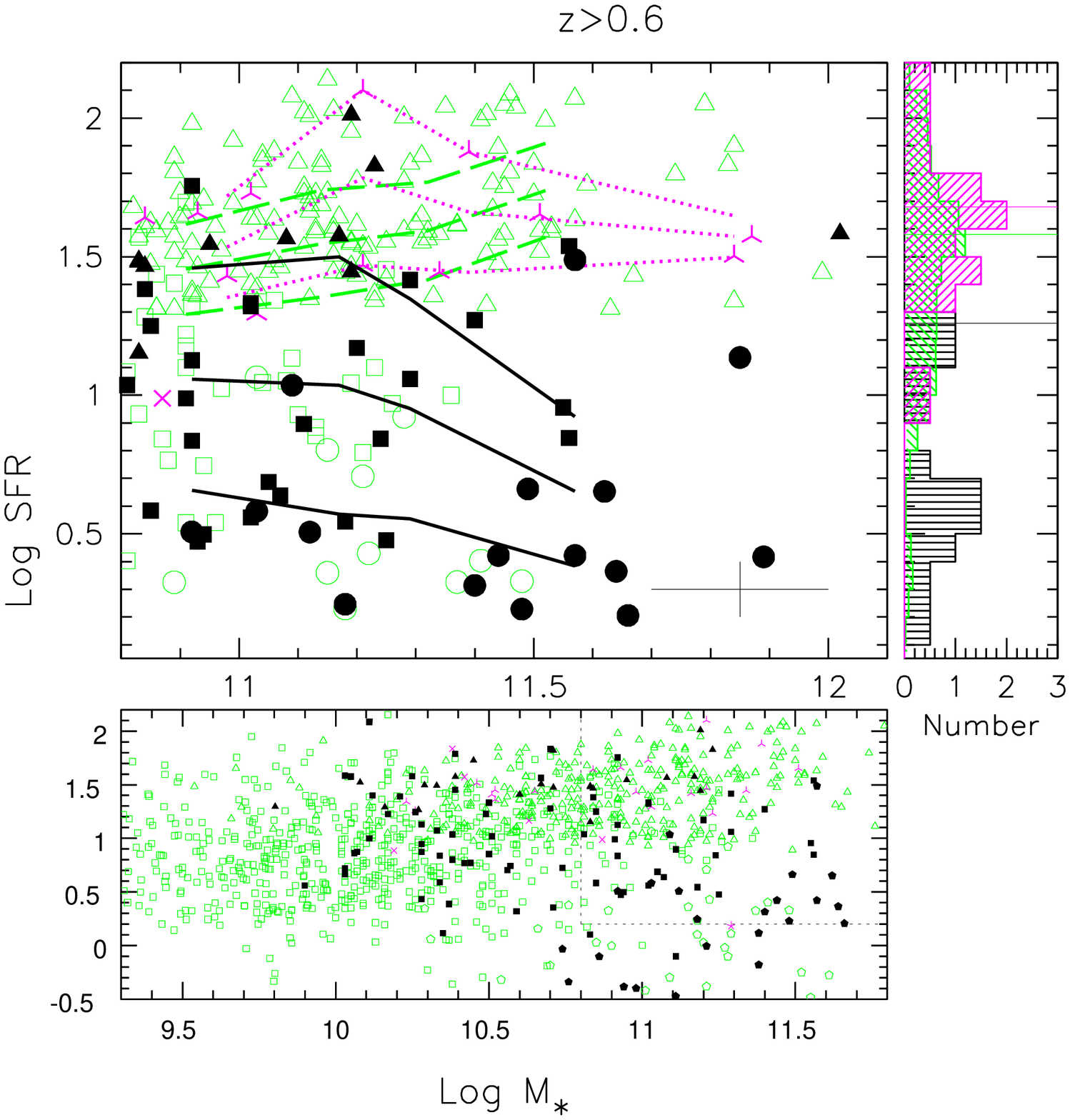}
\vspace*{2.2cm}
\end{minipage}
\hspace{1cm}
\caption{Same as Fig.~1, now comparing the SFR--M$_{\ast}$ relation of
cluster, group and field galaxies only at $z>0.6$.  {\bf Left panel}:
only 24$\mu m$+BEL galaxies.  {\bf Right
panel}: including REL galaxies. {\it Field}: green dotted
lines and empty symbols as in Figs.~1 and ~2. {\it Clusters}: black
solid lines and filled symbols as in Figs.~1 and ~2. {\it Groups}:
magenta lines and symbols, skeletal triangles are 24$\mu m$-detected
galaxies, while crosses are BEL galaxies. No REL galaxy is present in groups.
\label{CGF} }
\end{figure*}

With the aim of investigating whether cluster and group galaxies,
separately, differ from the field, we divide the sample
into clusters with $\sigma > 400 \rm \, km \, s^{-1}$ and groups with
$\sigma < 400 \rm \, km \, s^{-1}$
(Fig.~\ref{CGF}). We only consider the highest-z bin, as the lowest-z
bin has too few group galaxies to study any trend.

Figure \ref{CGF} shows that the SF in the cluster environment 
deviates from the field trend, while group galaxies seem to follow the
SFR-Mass relation of the field. The K-S test perfromed on mass-matched samples 
yields a 98\% (without red galaxies) and a 99.9\% (with red galaxies)
probability
that clusters have a different SFR distribution
from {\it both} groups and field.
In contrast, 
the test cannot reject the hypothesis
 that the groups and field have a similar distribution.

Our group data are not sufficient to draw firm conclusions, but, if
confirmed, our finding suggests that the group environment is not
influential for the link between SF activity and mass, therefore 
strangulation could not be relevant, at least in groups, and only
cluster-specific processes could be important.

Having removed the groups, the mean SFR in clusters at $z>0.6$
is 1.93$\pm$0.02 (without red) to 2.13$\pm$0.02 (with red) times 
lower than in the field.

\section{Conclusions}

This is the first attempt to establish whether the relation between
SF activity and galaxy mass depends on environment.  
We find that this relation in clusters is significantly different
from the field, at all redshifts when REL galaxies 
are included, and at least at $z>0.6$ for 24$\micron$+BEL galaxies.
Discriminating star-forming from AGN red galaxies will allow us
to quantify with higher precision the environmental effects.

The observed differences between  the SFRs in clusters and in the field can be
considered a lower limit to the real gap. In fact, we are surely
considering as cluster members also 
star-forming galaxies that are 
either in projection or just recently accreted by the
cluster and have not been affected yet by its influence. 

This result cannot arise from having severely
underestimated the dust correction to the $SFR_{[OII]}$. 
If we were to adjust the [OII] extinction to bring  
the field and clusters into agreement,
many of the [OII] detected sources should be detected at 24$\micron$,
and they are not.

Environmental
differences are detectable thanks to the low completeness limit in SFR
reached. 
With a higher limit,
the cluster and field relations would
appear compatible. Hence, any comparison of SFR and
masses in different environments is meaningful only when all data probe
down to low SFR levels.

We conclude that
there are significant differences between the SF
activity of star-forming galaxies of the same mass in different
environments. Clusters, in general, show a lower SF activity than the
field, not only because they have a pre-existing large population of
early-type galaxies passively evolving since high-z, but because
currently star-forming galaxies host an average lower SFR
than their field counterparts of similar mass.

The most straightforward interpretation is that there are
environmental effects suppressing SF in clusters.  Fast-acting
mechanisms would leave the SFR-Mass relation unchanged, while
processes with a longer timescale would affect it.  The most popular
long-timescale candidate is strangulation, that, if equally effective
in groups and clusters, would be ruled
out if the similarity we observe between groups and field will be
confirmed by larger studies.  
Even ram pressure stripping, which acts
on a short timescale \citep{bekki09}, may leave residual gas and
low SFRs.

As an alternative to environmental mechanisms, 
it is possible that other galaxy intrinsic properties
besides mass (e.g., the morphological
distribution) influence the 
SF history and vary systematically with environment.

In clusters, we are
observing a population of galaxies in transition 
from being blue star-forming to red passively
evolving, while such a population is much
less noticeable in the field and 
perhaps also in groups.

\acknowledgments 
We thank the anonymous referee for her/his useful remarks, we thank Kai Noeske and the AEGIS
collaboration for providing us their data and support.  
We acknowledge financial support
from ASI contract I/016/07/0.


\begin{thebibliography}{}
\bibitem[Bai et al.(2006)]{bai06} Bai, L., Rieke, G.~H., Rieke, M.~J., Hinz, J.~L., Kelly, D.~M., \& Blaylock, M.\ 2006, \apj, 639, 827 
\bibitem[Bai et al.(2007)]{bai07} Bai, L., et al.\ 2007, \apj, 664, 181 
\bibitem[Bai et al.(2009)]{bai09} Bai, L., Rieke, G.~H., Rieke, M.~J., Christlein, D., \& Zabludoff, A.~I.\ 2009, \apj, 693, 1840 
\bibitem[Balogh et al.(2004)]{balogh04a} Balogh, M., et al.\ 2004, \mnras, 348, 1355 
\bibitem[Balogh et al.(2004)]{balogh04b} Balogh, M.~L., Baldry, I.~K., Nichol, R., Miller, C., Bower, R., \& Glazebrook, K.\ 2004, \apjl, 615, L101 
\bibitem[Bekki(2009)]{bekki09} Bekki, K.\ 2009, \mnras, 399, 2221 
\bibitem[Bell et al.(2003)]{bell03} Bell, E.~F., McIntosh, D.~H., Katz, N., \& Weinberg, M.~D.\ 2003, \apjs, 149, 289 
\bibitem[Bell \& de Jong et al.(2001)]{bell01} Bell, E.~F., \& de Jong, R.~S.\ 2001, \apj, 550, 212
\bibitem[Borch et al.(2006)]{borch06} Borch, A., et al.\ 2006, \aap, 453, 869 
\bibitem[Brinchmann et al.(2004)]{brinchmann04} Brinchmann, J., Charlot, S., White, S.~D.~M., Tremonti, C., Kauffmann, G., Heckman, T., 
\& Brinkmann, J.\ 2004, \mnras, 351, 1151 
\bibitem[Cassata et al.(2007)]{cassata07} Cassata, P., et al.\ 2007, \apjs, 172, 270 
\bibitem[Cooper et al.(2008)]{cooper08} Cooper, M.~C., et al.\ 2008, \mnras, 383, 1058 
\bibitem[Daddi et al.(2007)]{daddi07} Daddi, E., et al.\ 2007, \apj, 670, 156 
\bibitem[Dale \& Helou(2002)]{dale02} Dale, D.~A., \& Helou, G.\ 2002, \apj, 576, 159 
\bibitem[Desai et al.(2007)]{desai07} Desai, V., et al.\ 2007, \apj, 660, 1151 
\bibitem[Elbaz et al.(2007)]{elbaz07} Elbaz, D., et al.\ 2007, \aap, 468, 33 
\bibitem[Feulner et al.(2005)]{feulner05} Feulner, G., Gabasch, 
A., Salvato, M., Drory, N., Hopp, U., \& Bender, R.\ 2005, \apjl, 633, L9 
\bibitem[Finn et al.(2009)]{finn09} Finn, R., et al.\ 2009, \apj, submitted
\bibitem[Font et al.(2008)]{font08} Font, A.~S., et al.\ 2008, \mnras, 389, 1619 
\bibitem[Fritz et al.(2009)]{fritz09} Fritz, J., et al.\ 2009, in preparation
\bibitem[Halliday et al.(2004)]{halliday04} Halliday, C., et al.\ 2004, \aap, 427, 397 
\bibitem[Kennicutt(1998)]{kennicutt98} Kennicutt, R.~C., Jr.\ 1998, \araa, 36, 189 
\bibitem[Kewley et al.(2004)]{kewley04} Kewley, L.~J., Geller, M.~J., \& Jansen, R.~A.\ 2004, \aj, 127, 2002 
\bibitem[Kere{\v s} et al.(2005)]{keres05} Kere{\v s}, D., Katz, N., Weinberg, D.~H., \& Dav{\'e}, R.\ 2005, \mnras, 363, 2 
\bibitem[Kroupa(2001)]{kroupa01} Kroupa, P.\ 2001, Dynamics of Star Clusters and the Milky Way, 228, 187 
\bibitem[Larson et al.(1980)]{larson80} Larson, R.~B., Tinsley, B.~M., \& Caldwell, C.~N.\ 1980, \apj, 237, 692 
\bibitem[McCarthy et al.(2008)]{mccarthy08} McCarthy, I.~G., Frenk, C.~S., Font, A.~S., Lacey, C.~G., Bower, R.~G., Mitchell, N.~L., Balogh, M.~L., \& Theuns, T.\ 2008, \mnras, 383, 593 
\bibitem[Milvang-Jensen et al.(2008)]{milvang-jensen08} Milvang-Jensen, B., et al.\ 2008, \aap, 482, 419 
\bibitem[Noeske et al.(2007a)]{noeske07a} Noeske, K.~G., et al.\ 2007, \apjl, 660, L43
\bibitem[Noeske et al.(2007b)]{noeske07b} Noeske, K.~G., et al.\ 2007, \apjl, 660, L47 
\bibitem[Pannella et al.(2009)]{pannella09} Pannella, M., et al.\ 2009, \apjl, 698, L116 
\bibitem[Pell\'o et al.(2009)]{pello09} Pell\'o, R., et al.\ 2009, \aap, submitted 
\bibitem[P{\'e}rez-Gonz{\'a}lez et al.(2005)]{perez05} P{\'e}rez-Gonz{\'a}lez, P.~G., et al.\ 2005, \apj, 630, 82 
\bibitem[Poggianti et al.(2006)]{poggianti06} Poggianti, B.~M., et al.\ 2006, \apj, 642, 188 
\bibitem[Poggianti et al.(2008)]{poggianti08} Poggianti, B.~M., et al.\ 2008, \apj, 684, 888 
\bibitem[Poggianti et al.(2009)]{poggianti09} Poggianti, B.~M., et al.\ 2009, \apj, 693, 112 
\bibitem[Rines et al.(2005)]{rines05} Rines, K., Geller, M.~J., Kurtz, M.~J., \& Diaferio, A.\ 2005, \aj, 130, 1482 
\bibitem[Salpeter(1955)]{salpeter55} Salpeter, E.~E.\ 1955, \apj, 121, 161 
\bibitem[S{\'a}nchez-Bl{\'a}zquez et  al.(2009)]{Sanchez09} S{\'a}nchez-Bl{\'a}zquez, P., et al.\ 2009, \aap, 499, 47 
\bibitem[Sud{\v z}ius et al.(1996)]{sudzius96} Sud{\v z}ius, J., Bobinas, V., \& Raudeliu, S.\ 1996, Memorie della Societa Astronomica Italiana, 67, 1079 
\bibitem[Yan et al.(2006)]{yan06} Yan, R., Newman, J.~A., Faber, S.~M., Konidaris, N., Koo, D., \& Davis, M.\ 2006, \apj, 648, 281 
\bibitem[Weinmann et al.(2006)]{weinmann06} Weinmann, S.~M., van den Bosch, F.~C., Yang, X., \& Mo, H.~J.\ 2006, \mnras, 366, 2 
\bibitem[White et al.(2005)]{white05} White, S.~D.~M., et al.\ 2005, \aap, 444, 365 
\bibitem[Willmer et al.(2006)]{willmer06} Willmer, C.~N.~A., et al.\ 2006, \apj, 647, 853 
\bibitem[Wolf et al.(2009)]{wolf09} Wolf, C., et al.\ 2009, \mnras, 393, 1302 
\bibitem[Zheng et al.(2007)]{zheng07} Zheng, X.~Z., Bell, 
E.~F., Papovich, C., Wolf, C., Meisenheimer, K., Rix, H.-W., Rieke, G.~H., 
\& Somerville, R.\ 2007, \apjl, 661, L41 
\end{thebibliography}
\end{document}